\documentclass[sigconf]{acmart}
\AtBeginDocument{%
  \providecommand\BibTeX{{%
    \normalfont B\kern-0.5em{\scshape i\kern-0.25em b}\kern-0.8em\TeX}}}

\copyrightyear{2024}
\acmYear{2024}
\setcopyright{acmlicensed}\acmConference[CHI '24]{Proceedings of the CHI Conference on Human Factors in Computing Systems}{May 11--16, 2024}{Honolulu, HI, USA}
\acmBooktitle{Proceedings of the CHI Conference on Human Factors in Computing Systems (CHI '24), May 11--16, 2024, Honolulu, HI, USA}

\usepackage{graphicx}
\usepackage{caption}
\usepackage{subcaption}

\begin{document}

\setcopyright{none}

\acmConference[1st HEAL Workshop at CHI Conference on Human Factors in Computing Systems]{}{May 12}{Honolulu, HI, USA}

\acmDOI{}

\acmISBN{}

\settopmatter{printacmref=false}
\title{AffirmativeAI: Towards LGBTQ+ Friendly Audit Frameworks for Large Language Models}

\author{Yinru Long}
\authornote{Equal contributions, alphabetical order on the last name}
\email{yinru.long@vanderbilt.edu}
\orcid{0000-0003-4830-9894}
\affiliation{%
  \institution{Psychology and Human Development\\Peabody College\\Vanderbilt University}
    \streetaddress{230 Appleton Pl \#5721}
  \city{Nashville}
  \state{TN}
  \country{USA}
  \postcode{37203}
}

\author{Zilin Ma}
\authornotemark[1]
\email{zilinma@g.harvard.edu}
\orcid{0000-0002-7259-9353}
\affiliation{%
  \institution{Intelligent Interactive Systems Group\\Harvard School of Engineering and Applied Sciences}
    \streetaddress{150 Western Ave.}
  \city{Allston}
  \state{MA}
  \country{USA}
  \postcode{02134}
}

\author{Yiyang Mei}
\authornotemark[1]
\email{yiyang.mei@emory.edu}
\orcid{0009-0001-2923-0729}
\affiliation{%
  \institution{Law School\\Emory University}
    \streetaddress{1301 Clifton Road}
  \city{Atlanta}
  \state{GA}
  \country{USA}
  \postcode{30322}
}

\author{Zhaoyuan Su}
\authornotemark[1]
\email{nick.su@uci.edu}
\orcid{0000-0002-5647-8439}
\affiliation{%
  \institution{Donald Bren School of Information and Computer Sciences\\University of California Irvine}
    \streetaddress{204 Aldrich Hall Irvine}
  \city{Irvine}
  \state{CA}
  \country{USA}
  \postcode{90293}
}

\renewcommand{\shortauthors}{Ma and Mei, et al.}




\begin{abstract}
    LGBTQ+ community face disproportionate mental health challenges, including higher rates of depression, anxiety, and suicidal ideation. Research has shown that LGBTQ+ people have been using large language model-based chatbots, such as ChatGPT, for their mental health needs. Despite the potential for immediate support and anonymity these chatbots offer, concerns regarding their capacity to provide empathetic, accurate, and affirming responses remain. In response to these challenges, we propose a framework for evaluating the affirmativeness of LLMs based on principles of affirmative therapy, emphasizing the need for attitudes, knowledge, and actions that support and validate LGBTQ+ experiences. We propose a combination of qualitative and quantitative analyses, hoping to establish benchmarks for ``Affirmative AI,'' ensuring that LLM-based chatbots can provide safe, supportive, and effective mental health support to LGBTQ+ individuals. We benchmark LLM affirmativeness not as a mental health solution for LGBTQ+ individuals or to claim it resolves their mental health issues, as we highlight the need to consider complex discrimination in the LGBTQ+ community when designing technological aids. Our goal is to evaluate LLMs for LGBTQ+ mental health support since many in the community already use them, aiming to identify potential harms of using general-purpose LLMs in this context.
\end{abstract}

\maketitle

\section{Mental Health Disparities Experienced by LGBTQ+ Population}

Members of the LGBTQ+ community are disproportionately affected by mental health issues, evidenced by elevated rates of depressive symptoms, self-harm, and suicidal ideation, in stark contrast to their heterosexual and cisgender peers~\cite{amos_mental_2020, irish_depression_2019,toomey_coping_2018, semlyen_sexual_2016, veale_enacted_2017, trevorproject2023}. The act of coming out, while a significant step in one's identity, often exacerbates these challenges, leading to an increase in depression, anxiety, and thoughts of suicide~\cite{ryan_family_2010, cox_stress-related_2010, hilton_family_2011, meyer_prejudice_2003}. Minority stress theory highlights how societal stigma, discrimination, and internalized negative perceptions compound the psychological struggles faced by LGBTQ+ individuals, fostering a deep-seated sense of alienation~\cite{meyer_minority_1995, cox_stress-related_2010, herdt_introduction_1989}.

Moreover, the dismissal of LGBTQ+ youths' struggles as simply "teenage angst" aggravates their sense of isolation and misunderstanding, potentially leading to severe outcomes like homelessness~\cite{ryan_family_2009, robinson_conditional_2018}. Despite the critical role of social support from family and friends in mitigating these stresses, LGBTQ+ individuals often perceive less familial support than their heterosexual and cisgender counterparts and face additional challenges in peer relationships~\cite{clark2023adolescents, saewyc2011research}. Given the heightened levels of minority stress and social support deficits, there's an urgent need for accessible and effective mental health services tailored for this marginalized group.

\section{Mental Health Chatbots and LGBTQ+ people}
Due to the scarcity of mental health services available, many LGBTQ+ people have turned to LLM-based chatbots for mental health support~\cite{ma2024evaluating}. Large Language Models (LLMs) enable natural, context-aware chatbots (e.g, ChatGPT) through extensive datasets and probabilistic word sequencing~\cite{kasneci_chatgpt_2023}. Some even claim that these chatbots can reflect the nuances in LGBTQ+-related topics~\cite{edwards_lgbtq-ai_2021}. These chatbots' adaptability is enhanced by fine-tuning, allowing them to specialize without the need for manual knowledge bases, and in-context learning for relevant responses~\cite{ziegler_fine-tuning_2020, dong2023survey, kasneci_chatgpt_2023}. When these chatbots are used in therapy, their capability can supposedly improve interactivity and therefore improve therapeutic adherence~\cite{fitzpatrick_delivering_2017}.

However, LLMs can produce unpredictable or harmful responses, particularly in private and sensitive areas like mental health, sometimes offering less empathetic feedback than human therapists and generating misleading ``hallucinated'' responses~\cite{vaswani-attention, Wang2021AnEO, liang2023holistic, lee_mathematical_2023}.

LLMs may also perpetuate biases due to non-diverse training data from sources with inherent imbalances, such as Reddit and Wikipedia, or policies marginalizing minority voices in datasets, like YouTube's demonetization of trans content~\cite{PewRedditUsers2016, hill_wikipedia_2013, alkhatib_street-level_2019}. This can lead to stereotypical biases in LLM outputs~\cite{kurita-etal-2019-measuring, sheng-etal-2019-woman, gehman_realtoxicityprompts_2020, stochastic-parrot}. Despite updates to reflect changing societal dynamics, the high computational cost of retraining limits the frequency of updates, risking the perpetuation of outdated stereotypes and biases~\cite{twyman_2017, polletta_contending_1998, stochastic-parrot}. Moreover, LLMs cannot fully comprehend LGBTQ+ experiences due to their lack of authentic human experience~\cite{edwards_lgbtq-ai_2021}. 

The research by Ma et al.~\cite{ma_understanding_2023, ma2024evaluating} shows that LGBTQ+ individuals value chatbots' immediate support and the convenience they offer, creating a confidential space for deeply personal discussions. They also help to build strong emotional connections between the users and the chatbots, which can be particularly helpful in developing social skills. This ease of use and the potential for emotional attachment may promote consistent engagement with therapeutic practices in mental health contexts, although it could also lead to an over-dependence on these digital tools.

Chatbots also serve an additional purpose by providing support that may be lacking in their real-life environments~\cite{ma2024evaluating}. LGBTQ+ people turn to these chatbots for advice on managing discrimination, for affirmation of their identities, and for practicing scenarios unique to the LGBTQ+ experience, such as coming out or navigating LGBTQ+ dating scenes. However, the study also highlights limitations. Chatbots might not fully grasp the complex emotional needs specific to LGBTQ+ individuals. Generalized, vague, and empty statements that focus on boilerplate solutions failed to meet LGBTQ+ people's complex needs. Moreover, the advice given by chatbots can sometimes be out of touch with evolving social norms, posing risks to users if taken at face value, especially in sensitive situations like coming out to unsupportive family members.

\section{LGBTQ+ Affirmative Framework}

Affirmative therapy is a type of psychotherapy used to validate and advocate for the needs of sexual and gender minority clients~\cite{hinrichs2017affirmative}. Rosati et al.~\cite{Rosati2022NonBinary} pointed out that the lack of \textit{affirmativeness} in mental health providers such as therapists could not only fail to establish a trustful therapeutic alliance but also have the potential to produce harm through microaggression and unfamiliarity of gender issues. In addition, other key clinical issues were recommended for therapists to consider while working with LGBTQ+ individuals include sexual and gender identity development, couple relationships and parenting, family roles, unique minority stressors such as religious conflict, discrimination, victimization, legal and workplace issues~\cite{pachankis_clinical_2004}. 

However, affirmativeness is hard to quantify. Only general guidelines such as training protocols for affirmative therapists currently exist. APA guidelines suggest that it is important to take into account of the intersectionality of one's sexuality, gender, and other demographic attributes~\cite{hinrichs2017affirmative, pachankis_clinical_2004}. In 2000, the American Psychological Association (APA) first provided guidelines for working with LGB clients, and APA has been dedicated to refining the guidelines over the years based on minority stress theory~\cite{meyer_minority_1995, meyer_prejudice_2003} and affirmative psychology~\cite{moradi_engaging_2018}. 

Moradi and Budge~\cite{moradi_engaging_2018} suggested that the conceptualization of LGBTQ+ affirmative therapy should apply to all clients instead of having to assess particular identities first. Given the rise of large language model-based chatbots and the increasing use of such chatbots for mental health purposes~\cite{ma2024evaluating}, it is important to consider adapting affirmative guidelines for therapists into the context of interacting with chatbots. This way, it would facilitate a generally more affirmative interaction between chatbots and human users without making assumptions about user identities.

Much literature has touched on the topic of general guidelines for improving LGBTQ+ affirmative attitude, knowledge, and therapeutic skills~\cite{pepping_affirmative_2018, pachankis_clinical_2004}. Some measures were developed to assess therapists' competency in attitude, knowledge, and skills, such as the sexual orientation counselor competency scale (SOCCS)~\cite{bidell2005sexualorientaion}. We argue that even though the skills sub-scale seemed to be specifically designed for human therapists with example questions like ``I have experience counseling gay male clients'', the knowledge and attitude subscales could be potentially adapted to test chatbot's reactions to such questions. Some example questions in the knowledge subscale include  ``being born a heterosexual person in this society carries with it certain advantages'', and ``I am aware some research indicates that LGB clients are more likely to be diagnosed with mental illnesses than are heterosexual clients''. Other example questions from the attitude subscale include ``the lifestyle of LGB client is unnatural or immoral'', and ``when it comes to homosexuality, I agree with the statement: `You should love the sinner but hate or condemn the sin' ''~\cite{bidell2005sexualorientaion}. Such questions could be useful in examining the attitudes and knowledge of chatbots in LGBTQ+-related topics. 

\section{Benchmarking LLMs with Affirmative Therapy Frameworks}

Given the challenges posed by LLM-based chatbots, it's essential to assess the affirmativeness of LLMs, especially since they can become a crucial support system for LGBTQ+ individuals.  When traditional mental health resources are out of reach due to various barriers like cost, accessibility, or personal constraints, LLM-based chatbots might be one of the few available options for support~\cite{ma2024evaluating}. Often, people might not initially use LLMs with mental health support in mind. However, without strict conversational guidelines, it's hard to prevent LGBTQ+ individuals from turning to general-purpose LLMs for mental health assistance. Considering the potential risks associated with using these models for mental wellness, we argue that all LLMs undergo thorough evaluation for their ability to provide affirmative and supportive responses.

Affirmativeness is hard to specify, however. Therefore, it is vital to have benchmarks that help define and measure what appropriate ``affirmativeness'' means for LGBTQ+ people. We ask the following questions:
\begin{itemize}
    \item How to quantify affirmativeness with affirmative therapy framework? 
    \item What characterizes an Affirmative AI?
\end{itemize}

Building on the conceptualization of LGBTQ+ affirmative therapy developed by Moradi and Budge~\cite{moradi_engaging_2018}, we propose 3 core principles including affirmative attitude, accurate knowledge, and appropriate action (3As) for building more affirmative chatbots. 

\begin{itemize}
    \item Attitude: Counteracting anti-LGBTQ+ attitudes and proactively enacting LGBTQ+ affirmative attitudes.
    \item Accurate Knowledge: Obtain accurate knowledge about LGBTQ+ people's experience.
    \item Action: Acknowledge the heterogeneity of LGBTQ+ people's interactions with the chatbot and integrate it into affirming users' challenges to power inequalities without making hetero-cisgenderism assumptions or making suggestions that lack the consideration of safety or context.
\end{itemize}

Approaches to test these core values can be building prompts of attitudes, approaches, and scenarios that reflect a wide range of LGBTQ+ experiences and challenges. These prompts can simulate interactions between LGBTQ+ individuals and the chatbot, focusing on various aspects of their identities, experiences, and the specific challenges they might face. For instance, prompts can include scenarios involving coming out, dealing with discrimination, exploring gender identity, and seeking support for relationship issues specific to LGBTQ+ individuals.

To evaluate the chatbot's performance against these prompts, a set of criteria based on the 3As framework can be developed:

\begin{itemize}
    \item Affirmative Attitude: The chatbot's responses should reflect a positive and accepting attitude towards LGBTQ+ identities and experiences. This includes using inclusive language, affirming the individual's identity and experiences, and avoiding any form of judgment or bias.

    \item Accurate Knowledge: The chatbot should demonstrate an understanding of LGBTQ+ issues, including awareness of the social, psychological, and health challenges faced by LGBTQ+ individuals. This entails providing information that is factual, up-to-date, and reflective of the diverse experiences within the LGBTQ+ community.

    \item Appropriate Action: The chatbot should offer responses that are sensitive to the individual's context and safety. This means suggesting resources, coping strategies, and advice that take into account the potential risks and challenges specific to LGBTQ+ individuals, including considerations for their physical, emotional, and social well-being.

\end{itemize}

To effectively evaluate a language model's (LLM) performance in providing support to LGBTQ+ individuals, one approach involves first collecting responses from therapists experienced in LGBTQ+ mental health to a set of predefined prompts. These expert responses serve as a benchmark for affirmativeness. Subsequently, the same prompts are presented to the LLM, and its responses are recorded. The evaluation process then involves a two-fold analysis: therapists review the LLM's responses to assess their alignment with best therapeutic practices, providing qualitative feedback. Simultaneously, a quantitative comparison is conducted between the LLM's responses and those of the therapists, focusing on metrics such as affirmativeness, empathy, relevance, and accuracy. This comprehensive evaluation method highlights areas where the LLM excels or falls short, guiding targeted improvements to enhance its effectiveness as a supportive tool for the LGBTQ+ community.

Importantly, by benchmarking LLMs' affirmativeness, we are not arguing to use LLMs primarily in mental health support. We also do not claim that an LLM that is affirmative can solve the LGBTQ+ people mental health issues. As Ma et al~\cite{ma2024evaluating} has pointed out, researchers should consider the complex discrimination in LGBTQ+ people's community when hoping to design technological solutions. Rather, we intend to evaluate LLMs for LGBTQ+ people mental health only because many LGBTQ+ people have already started to use LLMs for mental health support. By benchmarking LLMs with affirmativeness, we can anticipate the harms of LGBTQ+ people using general purposed LLMs for mental health support.

In conclusion, by aligning LLM-based chatbots with affirmative therapy principles and benchmarking their performance, we can work towards creating LLMs that offer supportive, informed, and affirming interactions for LGBTQ+ individuals seeking mental health support. 

\bibliographystyle{ACM-Reference-Format}
\bibliography{sample-base,kzg}

\appendix
\end{document}